\documentclass[oldversion,letter]{aa}

\usepackage{graphicx}
\usepackage{txfonts}
\begin{document}

   \title{The nature of the 2014--2015 dim state of  RW Aur revealed by X-ray, optical, and NIR observations}

\titlerunning{RW Aur during dim state}

   \author{P. C. Schneider\inst{1}
          \and
          H. M. G\"unther\inst{2}
          \and
          J. Robrade\inst{3}
          \and
          S. Facchini\inst{4}
          \and
          K. W. Hodapp\inst{5}
          \and
          C. F Manara\inst{1}
          \and
          V. Perdelwitz\inst{3}
          \and
          J. H. M. M. Schmitt\inst{3}
          \and
          S. Skinner\inst{6}
          \and
          S. J. Wolk\inst{7}          
          }

   \institute{European Space Research and Technology Centre (ESA/ESTEC), 
         Keplerlaan 1, 2201 AZ Noordwijk, The Netherlands,
               \email{cschneider@hs.uni-hamburg.de}
     \and  
              Massachusetts Institute of Technology,
            Kavli Institute for Astrophysics \& Space Research,
            77 Massachusetts Avenue, Cambridge, MA 02139, USA
      \and 
            Hamburger Sternwarte, Gojenbergsweg 112,
              21029 Hamburg, Germany
      \and  Institute of Astronomy, University of Cambridge,
            Madingley Road, Cambridge, CB3 0HA, United Kingdom 
      \and
           Institute for Astronomy, 640 N. A'ohoku Place, Hilo, HI 96720, USA   
      \and              
             University of Colorado, 392 UCB, Boulder, CO 80309, USA
      \and             
             Harvard-Smithsonian Center for Astrophysics,
             60 Garden Street, Cambridge, MA 02139, USA
         }

   \date{received; accepted}

  \abstract
   {The binary system RW Aur consists of two classical T~Tauri stars 
     (CTTSs). The primary recently underwent its second observed major dimming event
     ($\Delta V\,\sim2$\,mag). 
     We present new, resolved \emph{Chandra}  X-ray and UKIRT near-IR (NIR) data 
     as well as unresolved optical photometry obtained in the dim state to study
     the gas and dust content of the absorber causing the dimming.
     The X-ray data show that the absorbing column density increased from 
     $N_H<0.1\times10^{22}\,$cm$^{-2}$ during the bright
     state to $\approx2\times10^{22}\,$cm$^{-2}$ in the dim state.
     The brightness ratio between 
     dim and bright state at optical to NIR wavelengths shows only a moderate 
     wavelength dependence and the NIR 
     color-color diagram suggests no substantial reddening. 
     Taken together, this indicates gray      absorption by large grains ($\gtrsim1\,\mu$m) 
     with a dust mass column density of
     $\gtrsim2\times10^{-4}$\,g\,cm$^{-2}$.
     Comparison with $N_H$ shows that an absorber responsible for 
     the optical/NIR dimming and the X-ray absorption is compatible with the 
     ISM's gas-to-dust ratio, i.e., that grains grow in the disk 
     surface layers without largely altering the gas-to-dust ratio. Lastly, we
     discuss 
     a scenario in which a common mechanism can explain the long-lasting dimming
     in RW Aur and recently in AA Tau.
     }


   \maketitle
%

\section{Introduction}
Classical T~Tauri stars (CTTSs) are young stars accreting from their 
surrounding protoplanetary disk. The disk's gas-to-dust ratio is 
challenging to measure directly. Therefore, the ISM ratio of 100:1 is 
regularly assumed \citep[see review by][]{Williams_2011}. 
In the standard magnetospheric accretion model, 
matter is channeled along magnetic field lines from the inner edge
of the disk onto the stellar surface \citep{Bouvier_2007}. 

The young binary RW~Aurigae (distance 140\,pc) is
among the well-studied CTTS with dense photometric and spectroscopic
monitoring. 
It is composed of a jet-driving primary and
a secondary  separated by 1\farcs4; the masses of both stars are thought to be 
approximately solar. 
The system (A+B combined) shows complex periodic optical variability. 
Its magnitude and color change by a few tenths of mag on time 
scales of 2--5\,days \citep[e.g.,][]{Petrov_2001b, Petrov_2001a}.
Additional irregular variability of up to two magnitudes along the reddening vector is 
superimposed onto this periodic variability \citep[e.g.,][]{Grankin_2007,Petrov_2007}.
Observations of CO emission show that the primary's disk is truncated at $R<60\,$AU
and that diffuse emission extends well beyond the disk; both features 
are best explained by a close 
($\sim70$\,AU) approach of RW~Aur~B around A  
\citep{Cabrit_2006, Dai_2015}.
A half-year  dimming event in 2011 ($\Delta V\approx2$\,mag) was
attributed to a ``bridge'' of material related to the extended CO 
structure connecting both components \citep{Rodriguez_2013, Dai_2015}.

Towards the end of 2014, the system dimmed again by 2\,mag.
The similarity of both dimming events could suggest that the 
``extra absorber'' in 2014--2015 is again associated with the tidal stream.
However, \citet{Petrov_2015} found changes
in certain wind features and that the absorber covered only the star and its
immediate surroundings. They proposed that the occultation was related to an
interaction between the stellar wind and the inner disk, which is 
supported by enhanced M- and L-band fluxes \citep{Shenavrin_2015}.
In this case, there might be similarities with UXor events observed 
in more massive stars \citep[e.g.,][]{Grinin_1998}.
No substantial increase in column density of cold gas towards the dimmed star
was found \citep{Petrov_2015}.

RW~Aur has also been observed  by Chandra with ACIS-S for 54\,ks
during the bright state 
\citep[January 2013,][]{Skinner_2014}. 
The A-B pair was resolved and the data revealed that 
the secondary is more X-ray luminous than the primary 
and that it is moderately variable (factor of 1.5 within 60\,ks).

X-ray absorption is essentially independent of ionization, because only
the inner atomic shells contribute to X-ray absorption.
Typically, the ratio between X-ray absorption expressed as equivalent 
neutral hydrogen column density ($N_H$)
and optical extinction ($A_V$) is used as a measure of the 
absorber's gas-to-dust ratio. As RW~Aur is already well characterized
during the bright state, we use the increase in X-ray absorption 
to derive the gas and small grain content of the extra absorber,
where ``small'' means that the grains are not opaque in X-rays (a few $\mu$m) .

We present new X-ray
and optical/near-IR (NIR) data obtained during the
dim state in Sect.~2, derive the corresponding 
column densities in Sect.~3, and compare both measurements in
Sect.~4 to investigate the properties of the extra absorber and
the nature of the dimming event.

\vspace*{-0.2cm}
\section{Observations and data analysis}
The new X-ray, optical, and NIR observations obtained to characterize 
the dim state are summarized in Table~\ref{tab:overview}.
\vspace*{-0.1cm}

\subsection{\emph{Chandra} observation and data processing}
The new X-ray observation (exposure time: 35\,ks) was performed with the same setup as the
2013 observation of RW~Aur during the bright state, i.e.,
1/8 subarray readout was used to minimize the pileup 
of the X-ray bright B component
\citep[see ][]{Skinner_2014}.
Data reduction was performed using \texttt{ciao} version 4.6 \citep{ciao}.
Unless otherwise noted, the 0.3--10.0\,keV
energy range was used.

\begin{table}[t]
\setlength{\tabcolsep}{0.1cm}
\caption{Overview of the analyzed observations\label{tab:overview}}
\vspace*{-0.2cm}
\begin{tabular}{ l c c r}
\hline\hline
Observatory & ObsID & Detector & Date(s) \\
\hline
IRTF & 2015A023 & SpeX & 20-Mar-2015\\
\emph{Chandra} & 17644 & ACIS-S VFAINT & 16-Apr-2015\\
UKIRT & U/15A/D01 & UFTI &  18-Apr-2015\\
OLT & -- & -- & 09-20-Apr-2015 \\
AAVSO & -- & -- & various \\
\hline
\end{tabular}
\vspace*{-0.4cm}
\end{table}

\begin{figure}
\centering
\fbox{\includegraphics[width=0.4\textwidth]{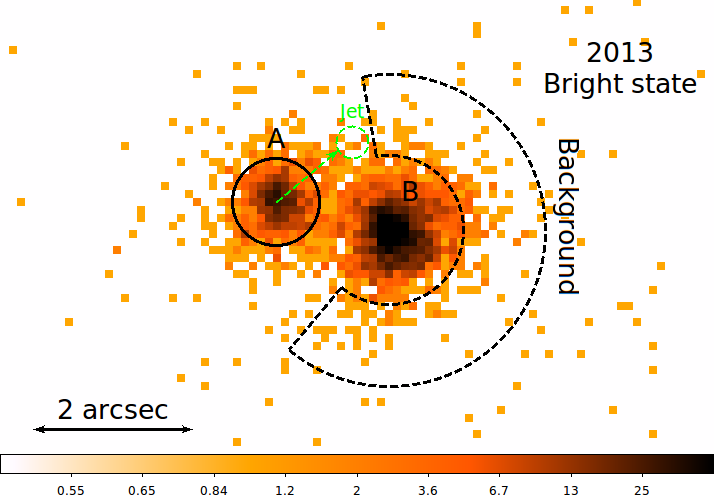}}
\fbox{\includegraphics[width=0.4\textwidth]{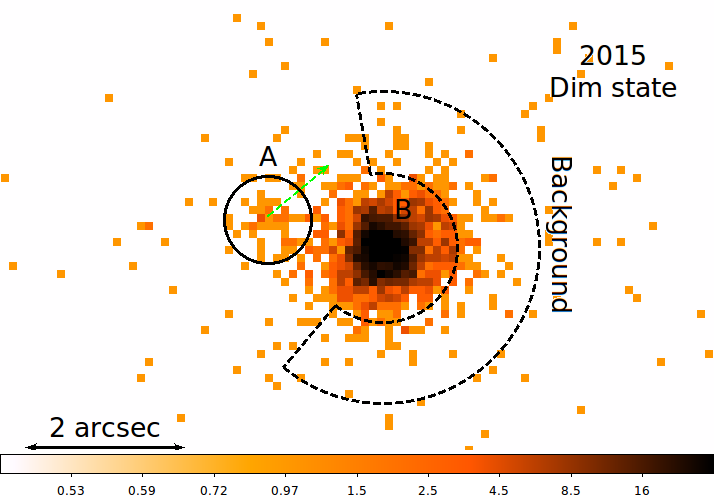}}

\vspace*{-5.35cm}\hspace*{-5cm}
\fbox{\includegraphics[width=0.12\textwidth]{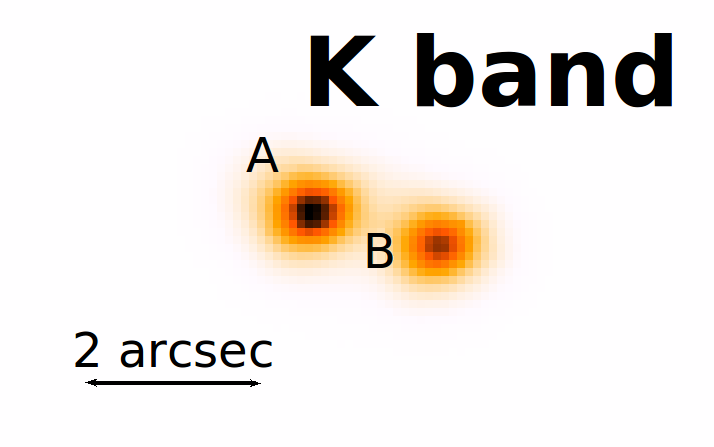}}
\vspace*{5cm}

\vspace*{-1cm}
\caption{X-ray images of RW~Aur taken during the bright state in 2013 (top) 
and during the dim state in 2015 (bottom); the color bars indicate
counts per pixel. The 
scaling reflects the difference in exposure time, i.e., similar count 
rates have similar colors. The inset shows the same sky area
in the  K band obtained two days after the X-ray observation. \label{fig:Ximage}}
\vspace*{-0.2cm}
\end{figure}

Figure~\ref{fig:Ximage} shows the X-ray images as well as
source and background extraction regions. 
The location of the 2013 jet emission (green arrow in Fig.\ref{fig:Ximage}) is outside 
our source extraction region and so  does not affect 
our analysis. The source region (radius: 0\farcs54, PSF 
fraction: 75\,\%) contains 51 photons and we estimate that 
PSF spillover by RW~Aur~B contributes 21.4 photons.
We checked that the exact parameters of the regions do not significantly 
influence the spectral data.

We modeled the stellar X-ray emission in \texttt{xspec}
using photoelectric absorption (\texttt{phabs}) and
\texttt{vapec} models with plasma abundances from
\citet{Skinner_2014} for consistency. 
The equivalent hydrogen column density 
depends on the assumed absorber abundances; we used \citet{angr}.
Lower metallicities as suggested by \citet{aspl} 
increase $N_H$ by about a factor of 1.5. 
To provide a robust estimate of the absorbing column 
density given the limited number of photons in 2015, 
we modeled the 2015 emission as
a scaled version of the 2013 two component model plus
extra absorption, i.e., we fixed the emission measure 
ratio ($EM_{hot}/EM_{cool}$) to values that accurately 
describe the 2013 \emph{Chandra} data. We also investigated 
several other models (e.g., free $EM_{hot}/EM_{cool}$
ratio, one component thermal plasma with free temperature) 
and find that the results are generally
consistent with this approach.

Here, we are exclusively interested in the evolution of
RW~Aur~A and  note that the average 
X-ray flux of the B component increased by a factor~1.4 
at similar spectral properties
(unabsorbed \hbox{$L_X = 1.9\times10^{30}$\,erg\,s$^{-1}$}) 
with respect to
2013 and that its count rate  smoothly decreases by a factor of~1.3
 during the exposure. X-ray emission 
from the RW~Aur~A jet as described by \citet{Skinner_2014} is not
significantly detected.

\subsection{Ground-based data}
Optical data in Johnson-Cousins filters were obtained with the 
Oskar-L\"uhning-Teleskop (OLT) at the Hamburger Sternwarte. Data 
reduction followed standard procedures. 
We also report 
AAVSO\footnote{Observations from the AAVSO International Database,
\texttt{http://www.aavso.org}} V magnitudes that densely sample the 
optical brightness evolution of the system. When near-simultaneous OLT 
data are available, both datasets are compatible.
Near-IR data were provided by IRTF and UKIRT. They were reduced using IRAF and
\texttt{Starlink}\footnote{\texttt{http://starlink.eao.hawaii.edu/starlink}},
respectively. 
The inset in Fig.~\ref{fig:Ximage} shows a UKIRT K-band image as an 
example of our NIR photometry.
Absolute photometry was obtained by calibration against the 2MASS magnitudes of
a check field and the standard star FS~12 \citep{Hawarden_2001} for the data from
20 March 2015 and 18 April 2015, respectively.

To estimate the brightness drop of RW~Aur~A, we compare 
the dim state values to its average bright state values,
which we estimate from the ROTOR data
\citep{Grankin_2007} transforming
their Johnson R  to Cousin~R using the \citet{Bessell_1979} 
relations.
For the NIR, we use the 2MASS magnitudes \citep{2MASS}
as the bright state reference.
The contribution of the B component to the optical/NIR flux is 
below 10\,\% during the bright state, but important during 
the dim state. For spatially unresolved observations, 
we subtract the average brightness of B calculated from 
resolved data to obtain the
magnitude of A (see Table~\ref{tab:magnitudes}).

\section{Results}
In the following, we show that the gas and dust absorption strongly  
increased between the bright and dim states.

\begin{figure}
\includegraphics[width=0.45\textwidth]{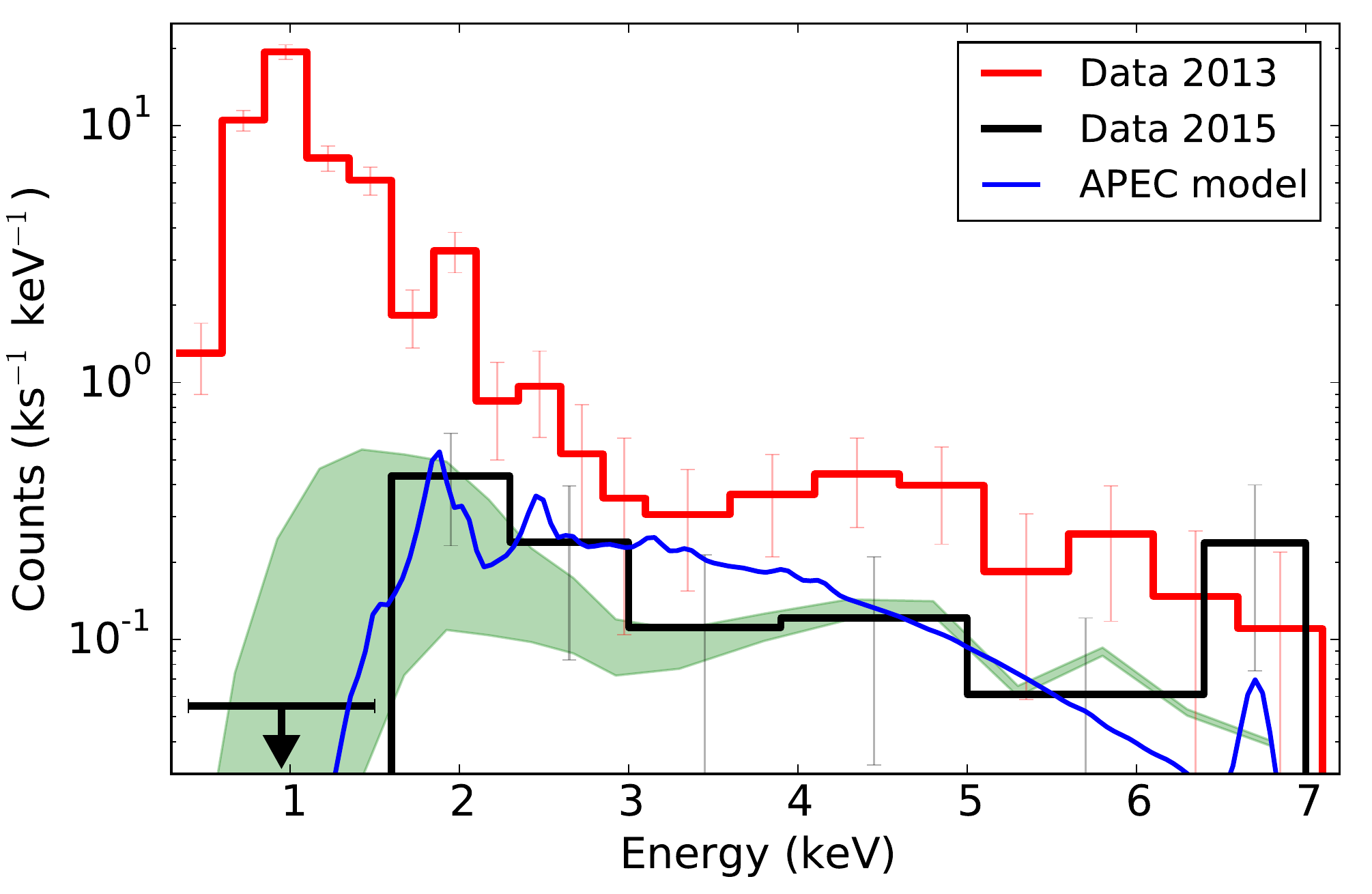}
\caption{X-ray spectra of RW~Aur~A with models. The APEC model
uses the parameters provided in Table~\ref{tab:xspec}. The green 
shaded area shows the 2013 data scaled to match at $E_{phot}>3$\,keV
including absorption; 
lower and upper bounds pertain to \hbox{$N_H^{extra}=10^{22}\,$cm$^{-2}$} 
and $4\times10^{22}\,$cm$^{-2}$, 
respectively. Data was binned for display purposes.
\label{fig:Xspec}}
\end{figure}

\subsection{X-ray data}

Figure~\ref{fig:Xspec} compares the X-ray spectra during the bright
and dim states. The observed (absorbed)
flux of the A component decreased by almost a factor of 
20 (see Table~\ref{tab:xspec}) mainly owing to the decrease at soft photon energies.
A significant source signal is only
recorded above 1.9\,keV. The number of photons at lower photon 
energies is compatible with a background fluctuation at
the 90\,\% confidence range. In addition, the observed flux above 2\,keV is reduced by 
about a factor of three.

The absorbing column density is rather well 
constrained because a hot component is needed to 
describe the high-energy tail of the observed spectrum 
($E_{phot}>2\,$keV) and emission from this component extends 
well into the soft part of the spectrum and drives the 
required absorbing column density.
The fit results, summarized in 
Table~\ref{tab:xspec}, show that the absorbing column density
increased by almost a factor of 30 to $N_H\approx2\times10^{22}$\,cm$^{-2}$
in 2015. The temperature of the hot component was fixed as its precise value
is  unconstrained \citep[already noted by][]{Skinner_2014}. However,
this does not affect the resulting $N_H$ significantly. 
The best fit column density corresponds to $5\times10^{-2}$\,g\,cm$^{-2}$ for a mean molecular 
weight of $\mu=1.4$
and to $A_V=13$\,mag for ISM-like absorption
\citep[$N_H = 1.8\times 10^{21}$\,cm$^{-2}$\,$A_V^{-1}$, ][]{Predehl_1995, Vuong_2003}.

\begin{figure}
\includegraphics[width=0.45\textwidth]{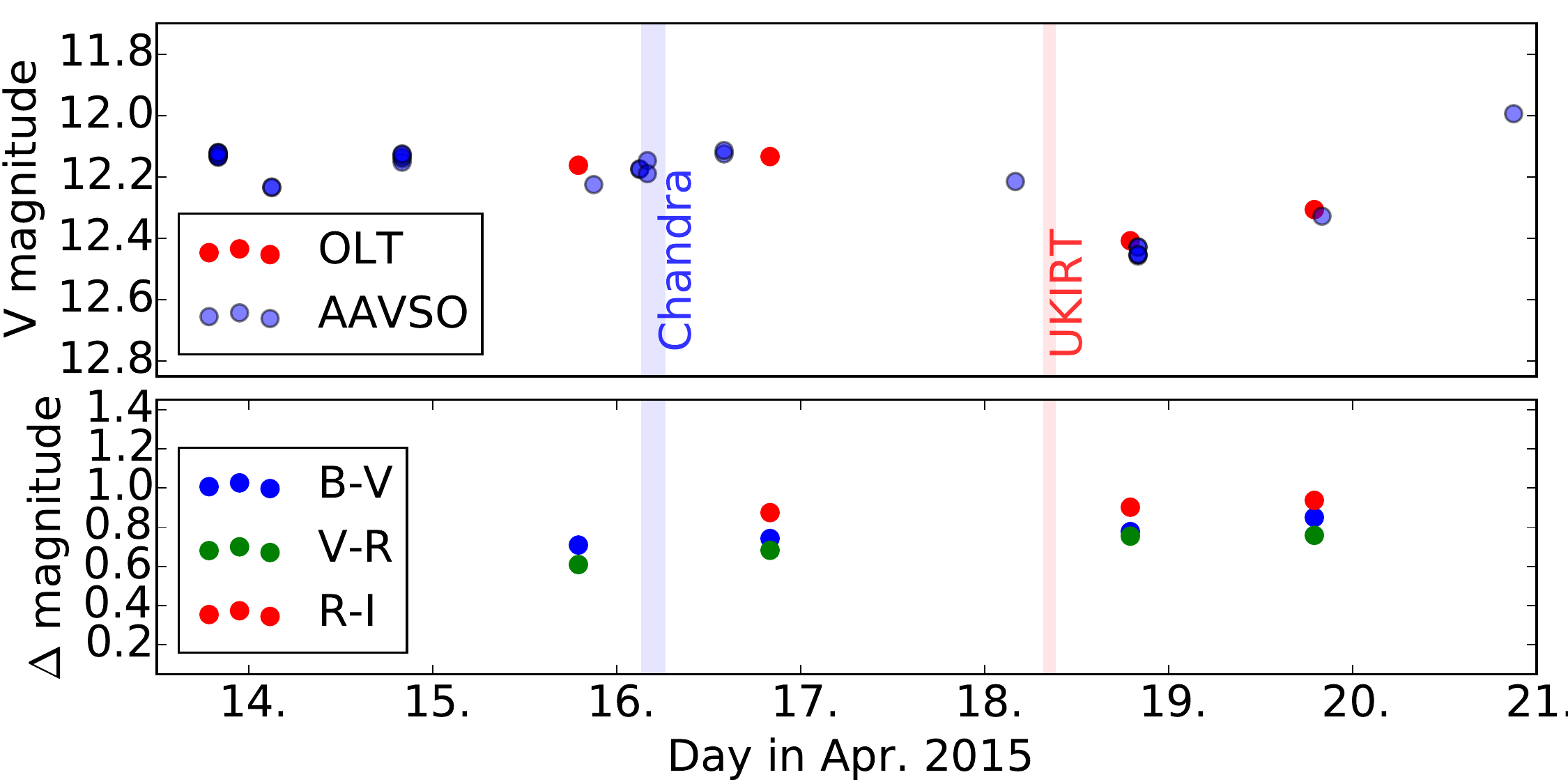}
\caption{Optical light curve of RW~Aur around the X-ray observation (spatially unresolved).
\label{fig:opt_lc}}
\end{figure}

\begin{table}[b]
\centering
\caption{X-ray properties of RW~Aur~A. Model with identical $EM_{hot}/EM_{cool}$ ratio in both epochs 
and k$T_{hot}$ fixed to 20\,keV. Errors indicate 90\,\% confidence ranges. 
Bin widths are 0.05 and 0.15\,keV for 2013 and 2015, respectively.
\label{tab:xspec}}
\begin{tabular}{ l c c c c c}
\hline\hline
Parameter & Value 2013 & Value 2015 & Unit \\
\hline\\[-0.2cm]
Rate (0.3 -- 2.0\,keV) & $12.3\pm0.4$&  $0.1\pm0.1$ & cts\,ks$^{-1}$\\
Rate (2.0 -- 10.0\,keV) & $2.0\pm0.2$ & $0.7\pm0.1$ & cts\,ks$^{-1}$\\[0.1cm]
$F_X$ (0.3-10\,keV): & $1.67\times10^{-13}$ & $4.7\times10^{-14}$ &  ergs\,cm$^{-2}$\,s$^{-1}$\\
$F_X$ (2.0-10\,keV): & $6.1\times10^{-14}$  & $2.8\times10^{-14}$  & ergs\,cm$^{-2}$\,s$^{-1}$\\[0.1cm]
$N_H$ & $0.08_{-0.06}^{+0.06}$ & $2.3_{-1.0}^{+1.6}$ & $10^{22}\,$cm$^{-2}$\\
k$T_{cool}$ & \multicolumn{2}{c}{$0.63_{-0.07}^{+0.12}$} & keV\\
k$T_{hot}$ & \multicolumn{2}{c}{$20.0$} & keV\\
$EM_{hot}/EM_{cool}$ & \multicolumn{2}{c}{$0.84\pm0.30$}\\
C-Statistic/bins &  180.30/221 & 61.45/65 & \\
\hline
\end{tabular}
\end{table}

\subsection{Optical and NIR data}
Figure~\ref{fig:opt_lc} shows the \hbox{V-band} light curve 
as well as the color evolution around the {\it Chandra} observation. 
Only minor variability is seen
around the X-ray observation (standard deviation in V
is 0.14\,mag on a ten-day time scale). Attributing this 
variability to the primary suggests intrinsic changes by only
0.2\,mag (color changes are $<0.1$\,mag), i.e., the optical 
properties are very similar during the
\emph{Chandra} and UKIRT observations. This strongly suggests
that the UKIRT data closely approximates the conditions during the 
X-ray observation. Compared to variability during the bright state,
these differences are small  so that we regard our 
X-ray, optical, and NIR datasets as effectively simultaneous.

\begin{figure}
\vspace*{-0.2cm}
\includegraphics[width=0.45\textwidth]{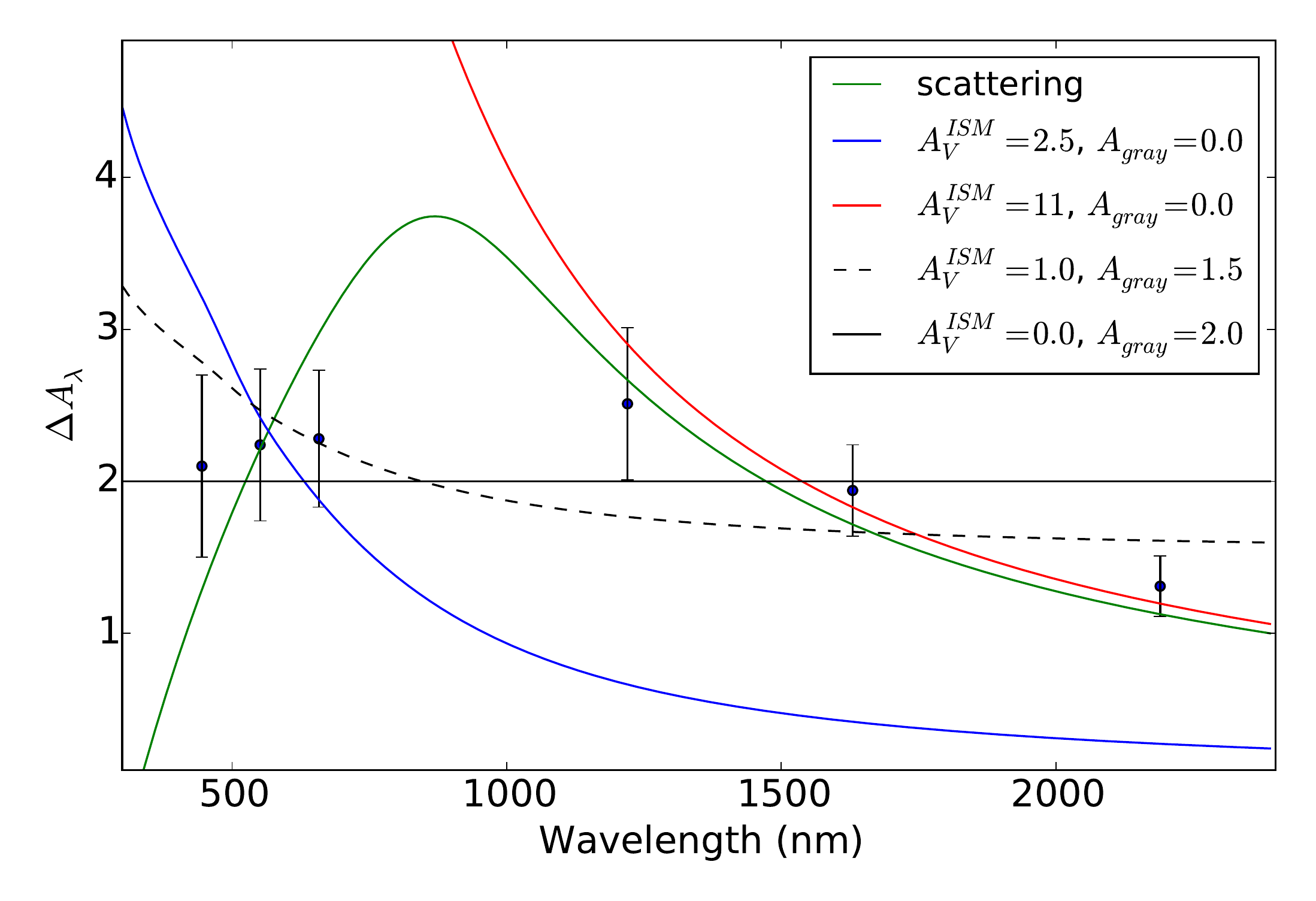}

\vspace*{-0.28cm}\includegraphics[width=0.45\textwidth]{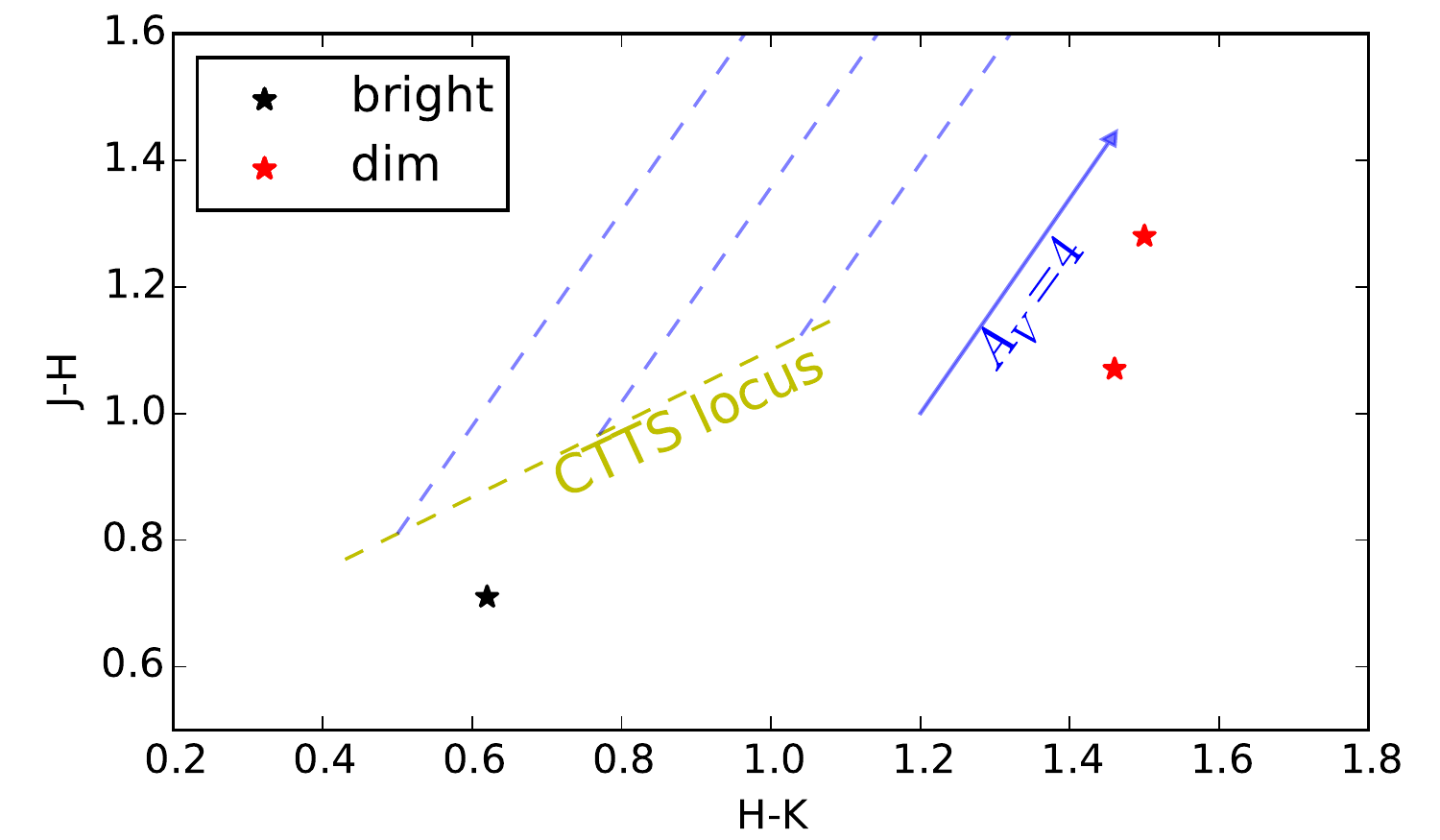}
\caption{{\bf Top}: Extinction by the extra absorber with different extinction curves.
ISM-like models have $A_{gray}=0$ and $R_V=3.1$.
The scattering model assumes ISM-like extinction with $A_V=10$\,mag, a scattering 
efficiency of 13\,\% in V, and no additional extinction along the scattering
path. Photometric bands are shown at their effective wavelengths. 
{\bf Bottom}: NIR color-color diagram.
\label{fig:extinction} \label{fig:color_color}}
\end{figure}

Figure~\ref{fig:extinction} shows the drop in brightness as a function
of wavelength with respect to the mean bright state.
The drop  at optical 
($\Delta \overline{BVR} \sim2.2$) and NIR 
($\Delta \overline{JHK} \sim1.9$) wavelengths is almost identical,
especially when considering that the bright state's reference magnitudes
are somewhat uncertain owing to intrinsic variability.
Therefore, ISM-like absorption, 
or minor variations thereof, are incompatible with the data (see blue and
red curves in Fig.~\ref{fig:extinction} top). 
Scattering of stellar photons in the circumstellar environment might 
contribute to the observed optical flux, which would erroneously 
indicate only little optical extinction possibly explaining why optical
and NIR fluxes show a similar drop. 
However, the NIR is essentially unaffected by scattering
because of its strong wavelength dependence 
($\sigma_{scat}\sim\lambda^{-4}$). Thus, substantial dust 
extinction should offset RW~Aur from the CTTS locus \citep{Meyer_1997}
along the reddening vector while the  
RW~Aur colors are below the CTTS
locus (Fig.~\ref{fig:color_color} bottom),
i.e., the NIR colors indicate no significant reddening. Without 
scattering, the evolution of the RW~Aur A optical/NIR magnitudes 
are almost wavelength independent, i.e., gray.
This requires an absorber consisting mainly of large grains 
and only a small amount of small grains, which is compatible
with the findings by \citet[][]{Antipin_2015} from their
optical data.

Gray absorption up to NIR wavelengths requires dust composed 
of grains with sizes $\gtrsim1\,\mu$m. 
The ratio between extinction and mass decreases
with increasing grain size. Therefore, assuming grains of $1\,\mu$m in size
 provides a lower limit on the dust mass.
These grains have absorption efficiencies around 2.5
\citep[see, e.g., ][]{Draine_1984} so that 
the mass attenuation coefficient is $10^4\,$cm$^{2}$\,g$^{-1}$
for dust densities of $\rho\approx2$\,g\,cm$^{-3}$ (e.g., carbonaceous grains, which
dominate ISM dust at these sizes). 
The required dust mass 
column density is about $2\times10^{-4}$\,g\,cm$^{-2}$ for gray 
extinction of 2\,mags.

\section{Interpretation and conclusions}

First, we assume the simplest scenario that reasonably explains our data: 
One single
absorber, which contains the gas that causes the X-ray absorption and 
the large dust grains that cause the gray optical/NIR extinction. 
The requirement of predominately large grains
is compatible with the 10/20\,$\mu$m features seen in some IR spectra of 
CTTSs, which also suggest large grains in the $\mu$m-size range
\citep[e.g.,][]{Oliveira_2010}. Combining X-ray
and dust absorption, we derive an 
 upper limit on the gas-to-dust ratio for this absorber of
250:1 with about a factor of two uncertainty (a factor of 1.7
in X-ray column density and we assume a factor of 1.5 in dust mass 
density $\rho$). Thus, one absorber with 
an ISM-like gas-to-dust ratio can explain the X-ray and 
optical data for processed dust, i.e., for dust with
sizes between about $1$ and $10\, \mu$m. This suggests that grains 
grow from their ISM distribution with a peak in the few $0.1\,\mu$m range to
sizes that are about ten times larger without strongly altering the 
gas-to-dust ratio. 
Without a substantial increase in low-ionization Na and K~{\sc i} 
absorption as suggested by 
\citet{Petrov_2015}, the strong increase in X-ray absorption indicates 
that the absorbing material is hot. This
suggests that the absorbing material is located
close to the star and might also be responsible
for the enhanced L- and M-band
fluxes measured by \citet{Shenavrin_2015}, e.g., the 
absorber might be a disk warp or an inner, dust-loaded 
wind as speculated by \citet{Petrov_2015}.

Second, we note that X-ray and dust extinction are not necessarily
cospatial. The gas absorption might be caused 
by dust-depleted accretion streams or winds launched from the disk rim as
in AA~Tau where the innermost region ($\sim0.1\,$au)
is strongly gas-enhanced \citep[$\Delta N_H\approx10^{22}$\,cm$^{-2}$; see][]{Schmitt_2007, Grosso_2007}.
In addition to this gas absorbing component, an opaque structure
that partly occults RW~Aur~A causes the gray extinction extending
from the NIR to the highest energies in our X-ray spectrum.
This opaque structure might be also located in the inner disk (0.1\,au) as the gas is, but
could also be  located slightly farther out at a few au -- like  the
absorber that causes the long-lasting dimming of AA~Tau 
\citep[][]{Bouvier_2013,Schneider_2015} -- or could even be part of the tidal stream.  
Depending on the actual 
location of the absorber, this likely requires a tilt between the inner and the
outer disk resolved by the CO observations given that the upper limit on 
the inclination of the RW~Aur~A disk is $i<60^\circ$, while that of AA~Tau is 
$i\approx75^\circ$. 
Such a scenario appears attractive, because it 
releases the constraint that the grain size distribution is missing
grains smaller than $\lesssim1\,\mu$m and 
provides a uniform explanation for AA~Tau-like CTTS and RW~Aur~A.

\begin{acknowledgements}
We thank B. Wilkes for granting the Chandra DDT observation and the referee, Prof. G. Gahm, for the careful
and constructive report.
PCS and CFM gratefully acknowledge an ESA Research Fellowship, HMG is supported by NASA-HST-GO-12315.01,
SJW NASA contract NAS8-03060 (Chandra), and SF by an STFC/Isaac Newton 
Trust studentship.
The results reported are based on observations made by the Chandra X-ray 
Observatory and by the United Kingdom Infrared Telescope (UKIRT) supported by NASA and operated 
under an agreement among the University of Hawaii, the University of Arizona, and Lockheed Martin 
Advanced Technology Center; operations are enabled through the cooperation of the Joint Astronomy Centre of the Science and Technology Facilities Council of the U.K.  and by UKIRT.
Some photometry was obtained at the Infrared Telescope Facility (IRTF), which is operated by
the University of Hawaii under contract NNH14CK55B with the National Aeronautics and Space Administration (NASA).
\end{acknowledgements}

\bibliographystyle{aa}
\bibliography{rw_aur}

\appendix

\section{RW Aur Photometry}
Table~\ref{tab:magnitudes} lists the optical/NIR magnitudes used in this study.

\begin{table*}
\centering
\caption{Optical and NIR magnitudes ordered by observing date. For unresolved observations, 
the magnitudes of the A component  were calculated by subtracting the average brightness of 
the B component estimated from the resolved observations with the error assumed as the
standard deviation ($B_B = 14.5\pm0.3$, $V_B=13.2\pm0.3$, $R_B=12.3\pm0.3$, $J_B = 10.06\pm0.3$, $H_B=9.14\pm0.1$, $K=8.61\pm0.04$ ). For resolved
data, the system's magnitudes are also provided by summing both components. \label{tab:magnitudes}}

\begin{centering}
\setlength{\tabcolsep}{0.1cm}
\begin{tabular}{ l c c c c c c c }
\hline\hline
Date & Band &  Resolved & System & RW Aur A & RW Aur B & Obs. & Ref\tablefootmark{a}\\
06-Sep-1986 & B & n & $11.18 \pm 0.48$ & $11.23\pm0.56$ & mean & ROTOR & (1) \\
\hspace*{0.2cm}-- 11-Oct-2013 & V & n & $10.45\pm0.40$  & $10.54\pm0.50$ & mean  \\
          & R$_J$ & n & $9.59\pm 0.31$ & \\
          & R$_C$ & n & $9.85\pm 0.31$ & $9.97\pm0.43$& mean \\
09-Nov-1994 & B & y & 10.99 &  $11.02\pm0.02$   & $14.89\pm0.10$  & HST & (2)\\
           & V & y & 10.45 & $10.51\pm0.02$   & $13.63\pm0.05$  & \\
           & R & y & 9.86 & $9.94\pm 0.03$   & $12.69\pm0.03$  & \\
06-Dec-1996 & K & & 6.83 & $7.06\pm0.17$    & $8.64\pm0.36$   & \\
27-Nov-1999 & J & n & $8.38 \pm 0.02$ & 8.64 & mean & 2MASS & (3) \\
           & H & n & $7.62\pm0.04$  & 7.93 & mean &  \\
           & \hspace*{0.8mm}K$_S$ & n &  $7.02 \pm 0.02$ & 7.31& mean \\
13/14-Dec-2014 & B & y & 13.62 & $14.50 \pm 0.06$ & $14.26 \pm 0.05$ & Caucasus  & (4) \\
 & V & y & 12.52 & $13.80 \pm 0.05$ & $12.92 \pm 0.03$ & \\
 & R$_c$ & y & 11.66 & $13.18 \pm 0.07$ & $11.97 \pm 0.07$ \\
20-Mar-2015 & J & y & 9.43 & $10.65 \pm 0.05$ & $9.85 \pm 0.04$ & UKIRT & (5) \\
           & H & y & 8.53 & $9.58 \pm 0.04$  &  $9.05 \pm 0.04$ \\
           & K & y & 7.57 & $8.12 \pm 0.03$  &  $8.57 \pm 0.03$\\
06-Apr-2015\tablefootmark{b} & V & n & $12.17\pm0.14$& 12.70 & mean & AAVSO & (5) \\
\hspace*{0.2cm} -- 26-Apr-2015 & R & n & $11.47\pm0.08$ & 12.15 & mean \\           
 09-Apr-2015\tablefootmark{b} & B & n & $13.02\pm0.15$ & 13.34 & mean & OLT & (5)\\
 \hspace*{0.2cm} -- 19-Apr-2015 & V & n & $12.25\pm0.11$ & 12.84 & mean \\
           & R & n & $11.55\pm0.07$ & 12.31 & mean \\
18-Apr-2015 & J & y & 9.90 & $11.15\pm0.05$   & $10.32\pm0.04$ & UKIRT & (5)\\ 
           & H & y & 8.76 & $9.87\pm0.03$    & $9.24\pm0.04$ \\
           & K & y & 7.74 & $8.37 \pm 0.04$  & $8.63\pm0.03$\\           
\hline
\end{tabular}\\
\end{centering}
\tablefoottext{a}{References as follows: (1) \citet{Grankin_2007}, (2) \citet{White_2001}, (3) \citet{Cutri_2003}, (4) \citet{Antipin_2015}, (5) This work. }
\tablefoottext{b}{Values pertain to mean and standard deviation during this time.}
\end{table*}

\end{document}